\documentclass[a4paper]{raa}
\usepackage{graphicx,times}
\usepackage{natbib}
\usepackage{amssymb,amsmath}
\usepackage{geometry}
\usepackage{lineno}

\def \kms {{ \rm km\;s$^{-1}$}}

\def \arcsec {$^{''}$}

\bibpunct{(}{)}{;}{a}{}{,}

\usepackage[pagebackref=true]{hyperref}
\hypersetup{pdftitle = The title of my PDF, pdfauthor = My name, pdfsubject= The subject, pdfkeywords = keyword1 keyword2 keyword3}
\hypersetup{colorlinks = true, linkcolor = green, anchorcolor = red, citecolor = blue, filecolor = red, pagecolor = red, urlcolor = red}

\begin{document}
%\linenumbers
   \title{Damping and power spectra of quasi-periodic intensity disturbances above a solar polar coronal hole$^*$
\footnotetext{\small $*$ Supported by the 973 project 2012CB825601 and National Natural Science Foundation of China.}
}
 \volnopage{ {\bf 2016} Vol.\ {\bf X} No. {\bf XX}, 000--000}
   \setcounter{page}{1}
   \author{Fangran Jiao\inst{1}, Lidong Xia\inst{1}, Zhenghua Huang\inst{1}, Bo Li\inst{1},  Hui Fu\inst{1}, Ding Yuan\inst{2}, Kalugodu Chandrashekhar\inst{1}
   }

   \institute{Shandong Provincial Key Laboratory of Optical Astronomy and Solar-Terrestrial Environment, Institute of Space Sciences, Shandong University, Weihai, 264209 Shandong, China; {\it xld@sdu.edu.cn}\\
        \and
            Jeremiah Horrocks Institute, University of Central Lancashire, Preston PR1 2HE, UK\\
\vs \no
   {\small Received 201x xxx xx; accepted 201x xxx xx}
}

\abstract{
We study intensity disturbances above a solar polar coronal hole seen in the AIA 171\,\AA\ and 193\,\AA\ passbands,
    aiming to provide more insights into their physical nature. The damping and power spectra of
    the intensity disturbances with frequencies from 0.07\,mHz to 10.5 mHz are investigated.
The damping of the intensity disturbances tends to be stronger at lower frequencies,
    and their damping behavior below 980\arcsec (for comparison, the limb is at 945\arcsec)
    is different from what happens above.
No significant difference is found between the damping of the intensity disturbances in the AIA 171\,\AA\
    and that in the AIA 193\,\AA.
 The indices of the power spectra of the intensity disturbances are found to be slightly smaller in the AIA 171\,\AA\
    than in the AIA 193\,\AA, but the difference is within one sigma deviation.
An additional enhanced component is present in the power spectra
in a period range of 8--40 minutes at lower heights. While the
power spectra of spicule is highly correlated with its associated
intensity disturbance, it suggests that the power spectra of the
intensity disturbances might be a mixture of spicules and wave
activities. We suggest that each intensity disturbance in the
polar coronal hole is possibly a series of independent slow
magnetoacoustic waves triggered by spicular activities.
\keywords{sun: corona --- sun: waves --- sun: turbulence ---
method: observational } }

\authorrunning{F. Jiao et al. }            %author_head in even pages
\titlerunning{Damping and power spectra of quasi-periodic disturbance}  % title_head in odd pages
\maketitle

\section{Introduction}           %% first-level sections will be auto-capitalized
\label{sect:intro}
Intensity disturbances in solar polar coronal holes have been discovered since \citet{1983SoPh...89...77W}, who reported a 10\% variation in the Mg\,{\sc x}\,625\,\AA\ line
    radiance in polar plumes.
The intensity disturbances tend to have a periodicity of 5--30 minutes\,\citep[e.g.][]{1983SoPh...89...77W, 1997ApJ...491L.111O, 1998ApJ...501L.217D},
    and propagate with a speed of 50--200\,\kms\,\citep[e.g.][among others]{1998ApJ...501L.217D,2009A&A...499L..29B,2011A&A...528L...4K,2015ApJ...809L..17J}.

Intensity disturbances are usually interpreted in terms of slow magnetoacoustic waves given that they propagate at roughly
    the sound speed\,\citep{1999ApJ...514..441O}.
\citet{2009A&A...499L..29B} found that the propagating speeds of the intensity disturbances increase from 75\,\kms\ at 0.6\,MK (Ne\,{\sc viii})
    to 125\,\kms\ at 1.3\,MK (Fe\,{\sc xii}), which is expected given the temperature dependence of sound speeds
    in the slow wave scenario.
Spectroscopic study using SUMER observations lent further support to this scenario\,\citep{2012A&A...546A..93G}.
The multi-passband data from SDO/AIA revealed that the propagation speed of intensity disturbances varies in different passpands\, \citep{2011A&A...528L...4K}.
Furthermore, the amplitude and damping length are smaller in hotter channels\,\citep{2011A&A...528L...4K,2012A&A...546A..50K}.
In the slow wave scenario, this temperature-dependence can be readily accounted for if electron heat conduction
    is the primary mechanism for the damping.
Recently, by analysing data from AIA 171\,\AA, 193\,\AA\ and 211\,\AA\ passbands,
    \citet{su2014a} discovered that the amplitude of intensity disturbances increases with height in the lower corona (0--9 Mm above the solar limb).
They found that the acoustic velocities inferred from the scale height highly correlate with the phase speeds of the intensity disturbances,
    and therefore concluded that the intensity disturbances in the lower corona are likely slow magnetoacoustic waves.
Furthermore, \citet{su2014b} statistically measured the phase speeds of intensity disturbances in polar regions observed by AIA 171\,\AA, 193\,\AA\ and 211\,\AA,
    and found that the speed ratios in different channels are consistent with the theoretical expectations for acoustic waves. By testing a 2.5D magnetohydrodynamic simulation to a flare loop, \citet{2015ApJ...813...33F} confirmed that the intensity disturbances following to chromospheric evaporation in the flare loop is slow mode waves.

An alternative interpretation for intensity disturbances is that they are quasi-periodic upflows, as was first suggested by  \citet{2010A&A...510L...2M}.
Using STEREO observations, \citet{2010A&A...510L...2M} found that high speed jets tend to occur quasi-periodically with a periodicity of 5--25 minutes,
    and travel along polar plumes with a speed of 135\,\kms\ and an apparent brightness enhancement of $\sim 5\%$.
A dedicated spectroscopic study using Hinode/EIS time-series observations revealed that repetitive high-velocity upflows,
    present as faint excess emission at $\sim$100\,\kms\ in blue wing of coronal lines,
    are correlated to the intensity disturbances in the corona\,\citep{2011ApJ...727L..37T}.

It is difficult to distinguish waves and flows in observations\,\citep{2015SoPh..290..399D} and it is not easy to separate different modes of waves since they can appear together in the same features\,\citep{2015A&A...581A..78Z}.
However, the wave and flow scenarios are not necessarily mutually exclusive.
Numerical studies by \citet{2012ApJ...754..111O} and \citet{2013ApJ...775L..23W} have shown that
    any upflow pulse at the coronal base results in propagating intensity disturbances guided by open flux tubes.
An observational evidence supporting this understanding was offered by \citet{2015ApJ...809L..17J},
    where they provided convincing evidence that spicular activities in the solar transition region
    are one of the important sources triggering intensity disturbances in the corona.
The connection between intensity disturbances in the corona and spicular activities in the lower atmosphere
    was recently confirmed by \citet{2015arXiv151107354S}.

While propagating, slow magnetoacoustic waves are expected to be dissipated by such mechanisms as
    compressive viscosity\,\citep{2000ApJ...533.1071O},
    electron thermal conduction\,\citep{2003A&A...408..755D} and mode coupling\,\citep{2004A&A...425..741D}.
They may also dissipate by developing into shocks\,\citep{2008ApJ...685.1286V}, given their amplitude growth
    in the presence of gravitational stratification\,\citep{2004A&A...415..705D}.
Hence, investigations into the damping of intensity disturbances can help understand whether the intensity disturbances are slow magnetoacoustic waves.
By analyzing observations from SDO/AIA, \citet{2014ApJ...789..118K} obtained the frequency-dependence of the damping length
    of the intensity disturbances observed in a polar coronal hole,
   and found that the damping length tends to increase rather than decreasing with frequency.
This discovery seems to be inconsistent with linear wave theory. However, the same analysis in on-disk plume-like structures\,\citep{2014ApJ...789..118K} shows the opposite behaviour.
\citet{2014A&A...568A..96G} also studied intensity disturbances in a solar polar coronal hole observed by the AIA 171\,\AA\ channel,
     and found that low-frequency waves propagate further compared to high-frequency waves.
Within a plume structure, he also found that the damping of intensity disturbances in the polar coronal hole
     follows two distinct profiles at two different height ranges (between 0--10 Mm and 10--70 Mm).

A power spectrum describes the power distribution in the frequency space.
It can be used to reveal whether any frequency is preferred in a broadband spectrum.
Studies of power spectrum are helpful to tell what happens in the solar atmosphere\,\citep{2015ApJ...798....1I}.
In a polar coronal hole observed by the AIA 171\,\AA\ channel,
    \citet{2014A&A...568A..96G} deduced power spectra of intensity disturbances at six individual points.
The spectra were found to follow Kolmogorov's law (power index$=$5/3),
    suggesting a possible connection between intensity disturbances and turbulence.

In this work, we examine intensity disturbances above a solar polar coronal hole in substantial detail, aiming to
    shed more light on the physical nature of intensity disturbances.
To this end, we examine the damping and power spectra of intensity disturbances as identified in two different AIA channels.
In the following, we describe the observations and methods of data analysis in Section\,\ref{sect:obs},
    present our results and discussion in Section\,\ref{sect:res},
    and give a summary in Section\,\ref{sect:sum}.

\begin{figure*}[!ht]
\includegraphics[width=\textwidth,clip,trim=0cm 0.3cm 0cm 0cm]{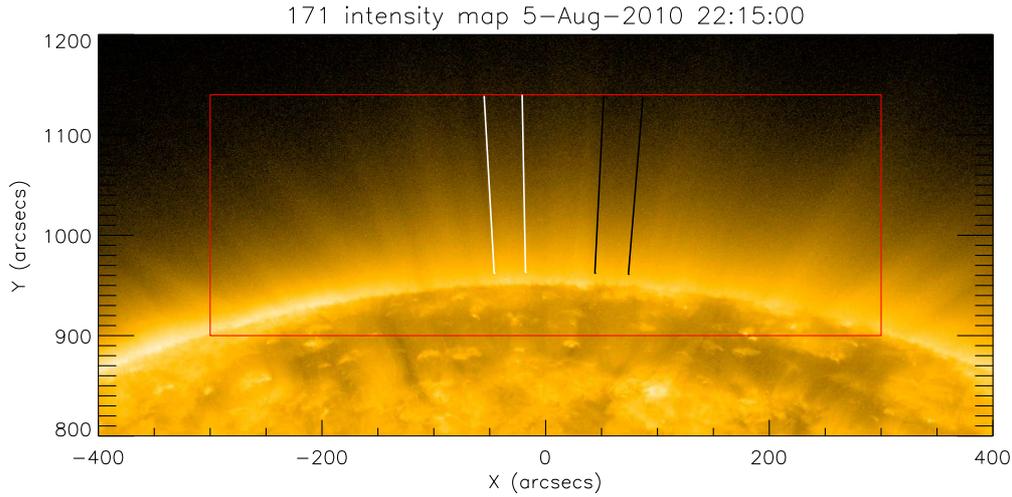}
\caption{
The studied polar coronal hole region (outlined by red lines) seen in AIA 171\,\AA.
The region between two white lines outline a region filled by plumes.
The region between two black lines mark the inter-plume region that is relative dark.
The intensity is shown on a log scale. \label{fig1}}
\end{figure*}

\section{Observations and data analysis}
\label{sect:obs}
The data used in this study were obtained from 21:50\,UT, 5 August 2010 to 01:50\,UT, 6 August 2010 by AIA\,\citep{2012SoPh..275...17L},
    with a spatial resolution of 1.2\arcsec and a cadence of 12\,s.
We aim to investigate intensity disturbances in the north polar region with solar X between $-$300\arcsec and 300\arcsec\ where a coronal hole is clearly seen (Figure\,\ref{fig1}).
The solar limb at X=0 is located at Y=945\arcsec.
Since intensity disturbances are a coronal phenomenon, we investigate the damping and power spectra of intensity disturbances observed by the two channels at coronal temperatures,
    namely  171\,\AA\ with response function peaking at 0.8 MK
    and 193\,\AA\ with response function peaking at 1.2 MK.
The fast Fourier transform (FFT) routine in the IDL library was used to
    extract the evolution of coronal features in the frequency space.

Figure~\ref{fig1} presents the three regions that we analyze:
   the entire coronal hole (the red box, 301 pixels at any given height above the limb),
   a plume region bordered by the white lines (18 pixels at any given height above the limb),
   and an inter-plume region sandwiched between the black lines (19 pixels at any given height above the limb) in Fig.\,\ref{fig1}.
Averaging the signals from the entire coronal hole substantially enhances the signal-to-noise ratio,
   thereby enabling one to examine the overall behavior of the intensity disturbances in this region.
On the other hand, distinguishing between a plume and an inter-plume region allows one to
   tell whether the physical properties of the intensity disturbances depend on the magnetic environment.

\begin{figure*}[!ht]
\includegraphics[width=\textwidth,clip,trim=1.5cm 3cm 0.6cm 0cm]{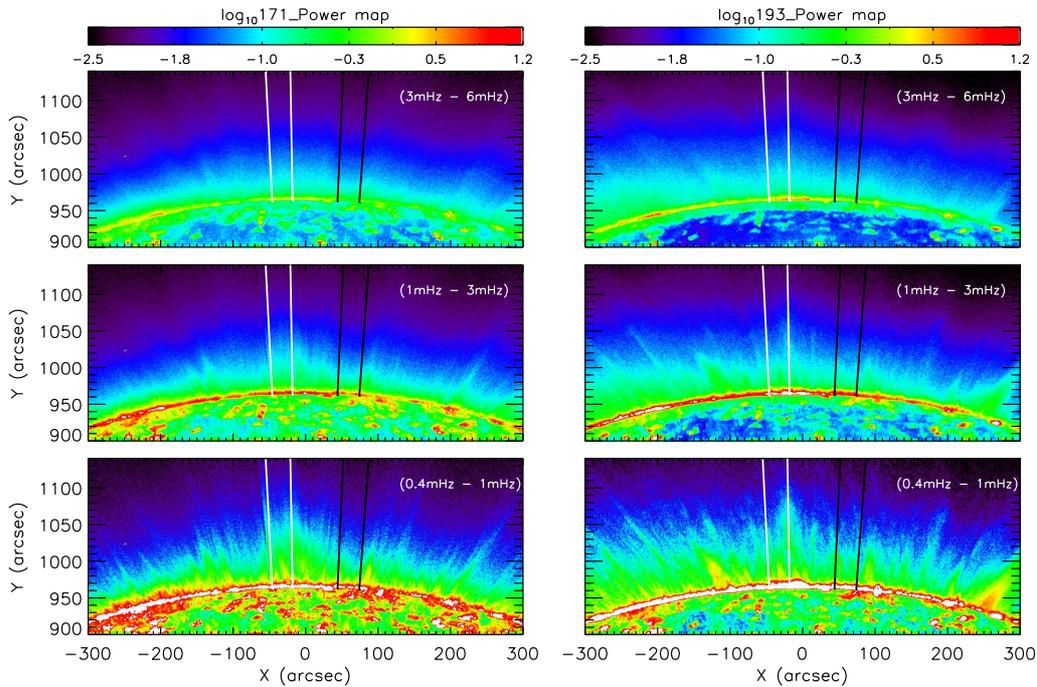}
\caption{Power map of the polar coronal hole at 3--6\,mHz (top), 1--3\,mHz (middle) and 0.4--1\,mHz (bottom).
The left (right) column displays the power map deduced from the AIA 171\,\AA\ (193\,\AA) emissions.
The plume and inter-plume regions are denoted by the white and black lines, respectively.
Here the Fourier power is shown on a log scale.}
\label{fig2}
\end{figure*}

\begin{figure*}[!ht]
\includegraphics[width=\textwidth,clip,trim=1.5cm 0.3cm 0.6cm 0cm]{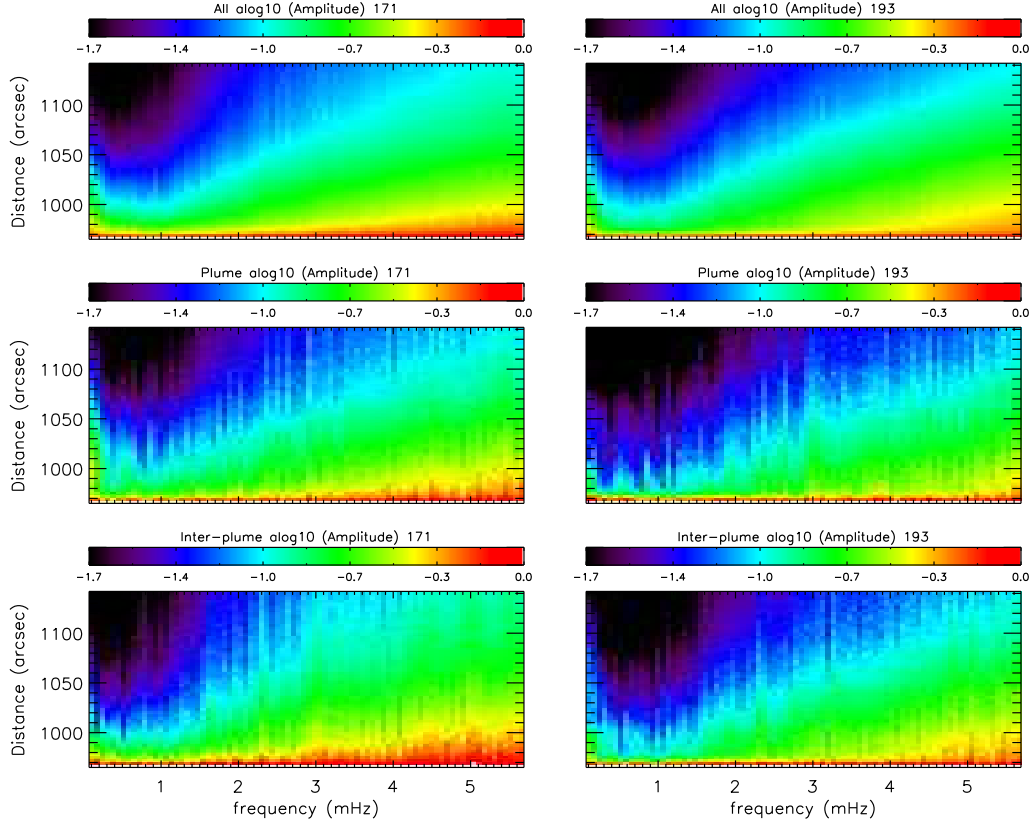}
\caption{The variation of normalised amplitude along the solar radial direction above the limb averaging from three different regions (top row: the whole coronal hole, middle row: plume region, bottom row: inter-plume region, see Section\,\ref{sect:obs} and Fig.\,\ref{fig1} for details of the locations of these regions). The shown amplitude at any particular frequency is normalised by that at the same frequency at the radius of 965\arcsec. The intensity disturbances in AIA 171\,\AA\ (left column) and 193\,\AA\ (right column) are displayed. \label{fig3}}
\end{figure*}

\begin{figure*}[!ht]
\includegraphics[width=0.8\textwidth,clip,trim=0cm 1cm 0.6cm 0cm]{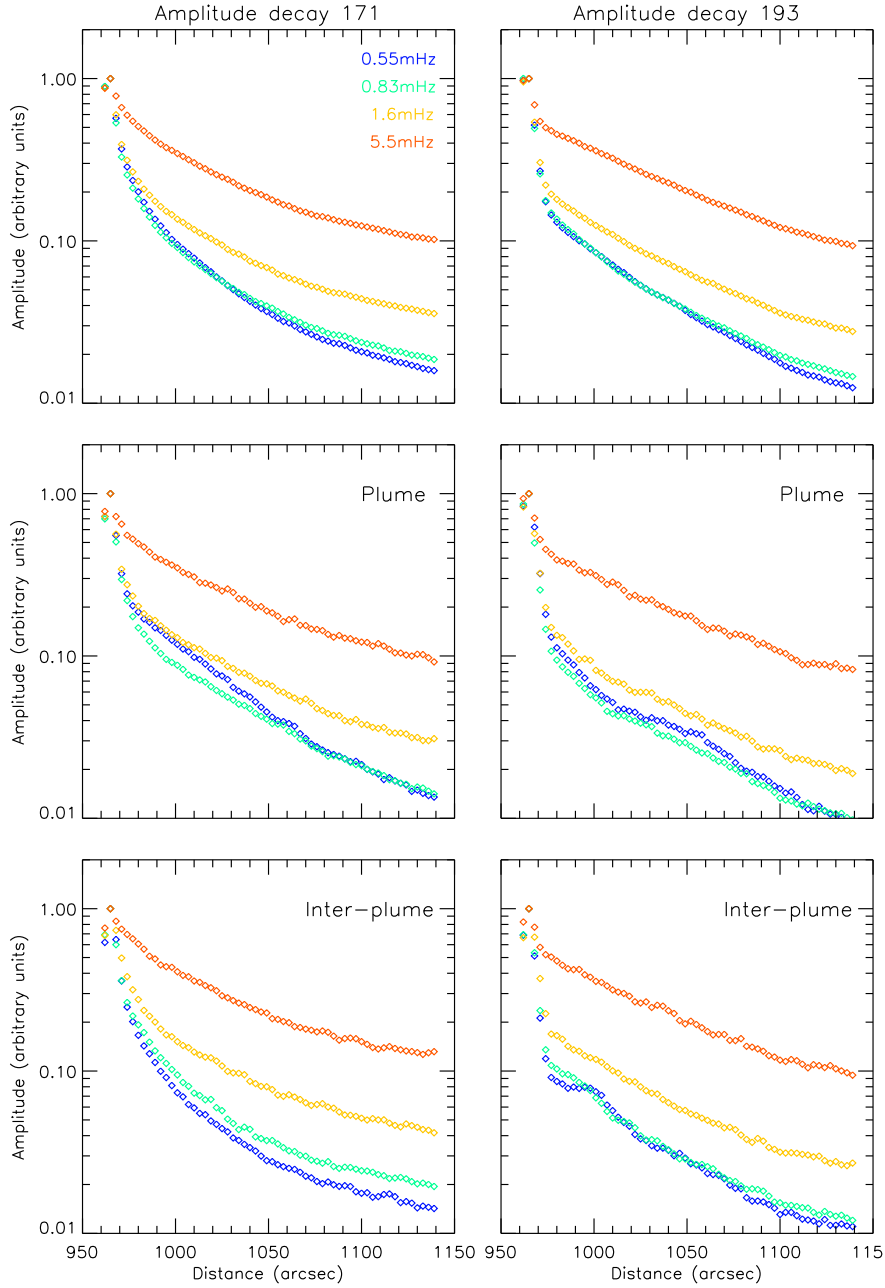}
\caption{ Damping of the intensity disturbances in the polar coronal hole. The results from the AIA 171\,\AA\ emission are shown in the left column and the 193\,\AA\ in the right column. Top row: coronal hole average, middle row: plume region and bottom row: inter-plume region. \label{fig4}}
\end{figure*}

\begin{figure*}[!ht]
\includegraphics[width=0.8\textwidth,clip,trim=0cm 1cm 0.6cm 0cm]{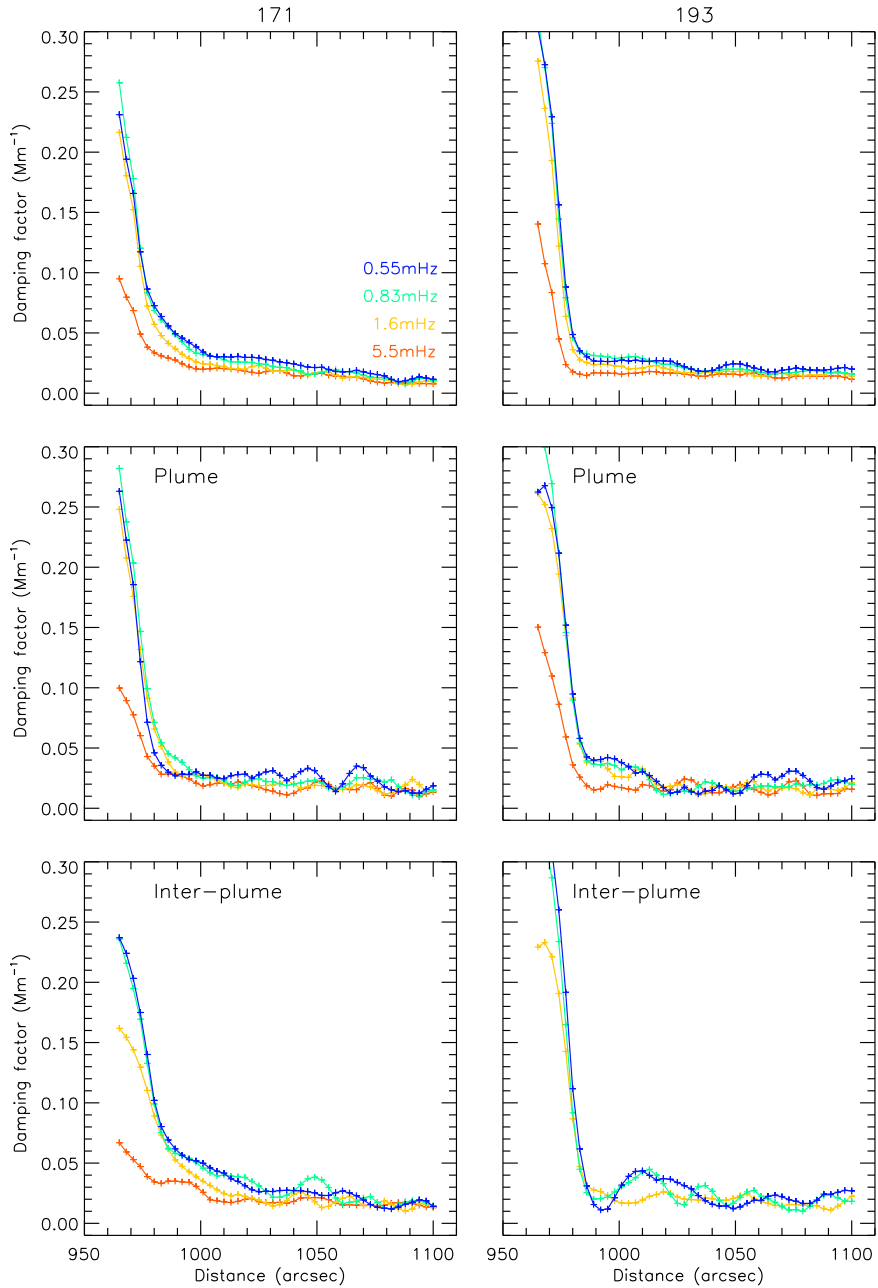}
\caption{Local damping factor deduced from the log-normal plots shown in Fig.\,\ref{fig4}.\label{fig5}}
\end{figure*}

\begin{figure*}[!ht]
\includegraphics[width=\textwidth,clip,trim=0cm 0.5cm 0.6cm 0cm]{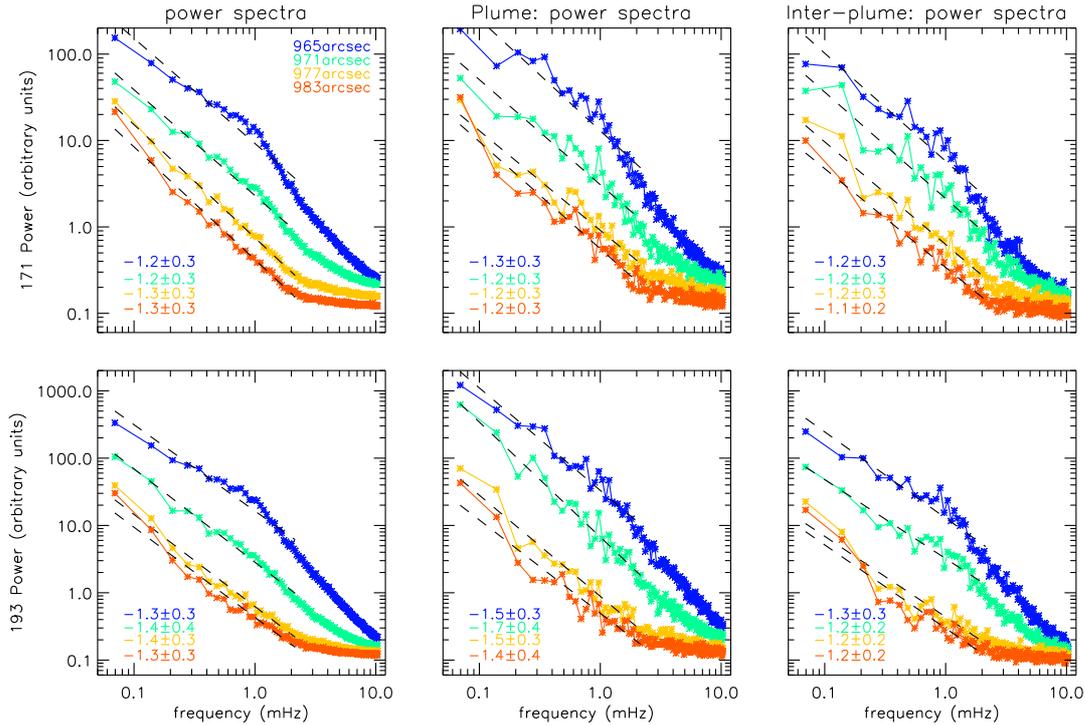}
\caption{Power spectra of the intensity disturbances in the polar coronal hole. The spectra are obtained from four different heights (blue: 965\arcsec, green: 971\arcsec, orange: 977\arcsec, red: 983\arcsec). The left row is the spectra from AIA 171\,\AA\ and the bottom one is from AIA 193\,\AA. The spectra in the left column are from coronal hole average, the middle column are from plume region and the right column are from inter-plume region.
The over-plotted dashed lines are fittings to the power-law spectra in the form $P(f)\propto f^{\alpha}$.
The errors of the power indices $\alpha$ are also given (see details in the main text)}.\label{fig6}
\end{figure*}

\begin{figure*}[!ht]
\includegraphics[width=\textwidth,clip,trim=4cm 1.8cm 0.5cm 3cm]{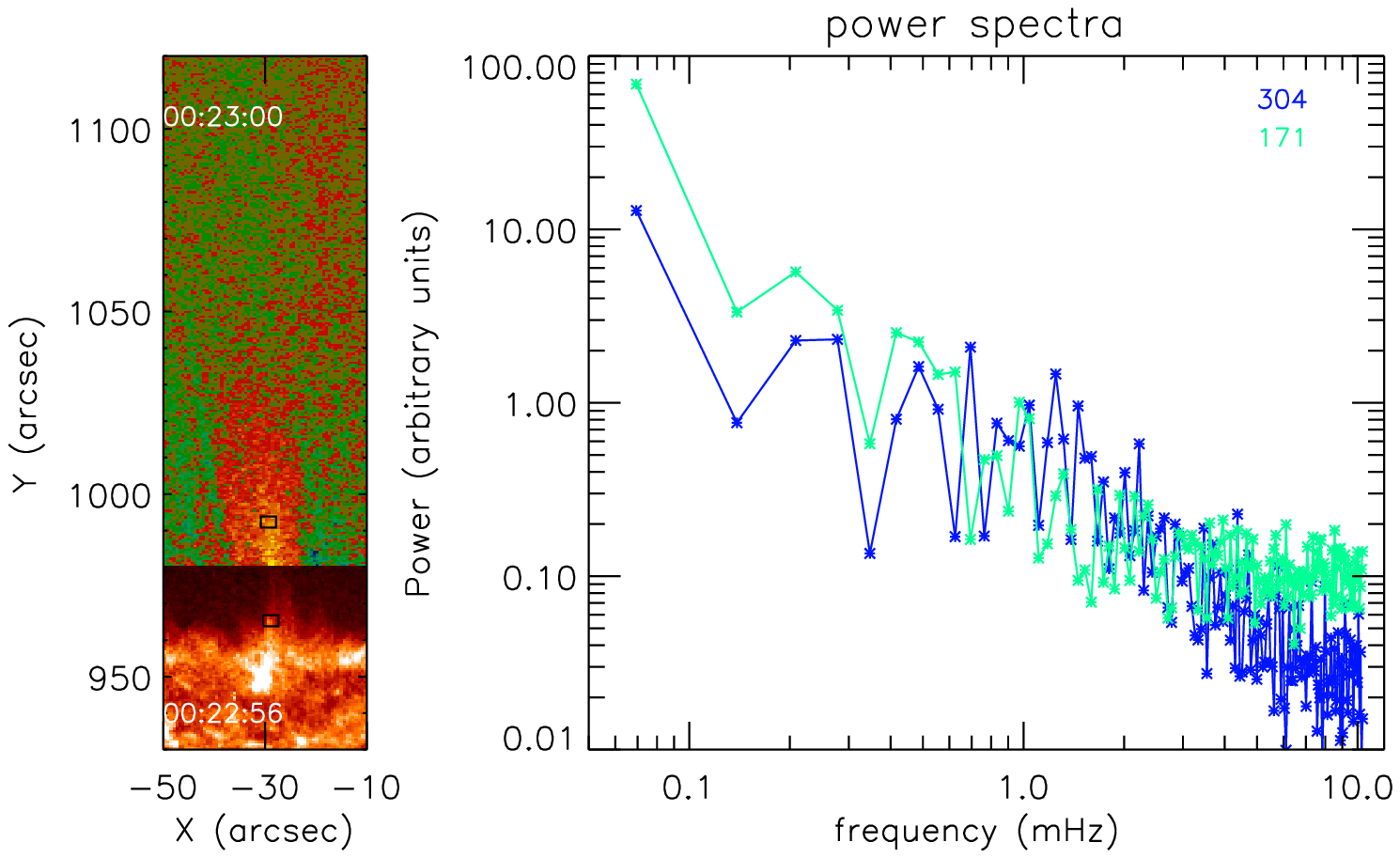}
\caption{Power spectra (right panel) of a spicule seen in AIA 304\,\AA\ (blue) and its associated intensity disturbance seen in AIA 171\,\AA\ (green). The regions to calculate the power spectra are shown in the left panel marked as black boxes, where the top one in the intensity disturbance seen in AIA 171\,\AA\ and the bottom one in the associated spicule seen in AIA 304\,\AA. \label{fig7}}
\end{figure*}

\section{Results}
\label{sect:res}

Fig.\,\ref{fig2} shows the power distribution in the polar coronal hole region
   in the frequency ranges 3--6\,mHz (the top row), 1--3\,mHz (middle) and 0.4--1\,mHz (bottom)
   seen in the AIA 171\,\AA\,(the left column) and 193\,\AA (right).
At low frequencies (0.4--1 mHz), the power images in both AIA 171\,\AA\ and 193\,\AA\ above the limb
   show many distinguishable radial structures filling the whole region.
Such radial structures are so-called intensity disturbances.
At higher frequencies, the intensity disturbances are mixed with each other and thus individual intensity disturbances are difficult to distinguish
   (see the top and middle panels in Fig.\,\ref{fig1}).
Comparing the left and right columns, one sees that
   the intensity disturbances as seen in AIA 171\,\AA\ are similar to those in AIA193\,\AA.
In line with\,\citet{2012A&A...546A..50K} and \citet{2014A&A...568A..96G},
   the power of the intensity disturbances decreases with altitude, suggesting that intensity disturbances are spatially damped.
In the next section, we provide a detailed examination on this damping together with
   its dependence on frequency and temperature.

\subsection{Damping of the intensity disturbances}
Figure~\ref{fig3} shows the amplitude distribution as a function of frequency
    and distance.
Because the intensity disturbances at a given height are much stronger at lower than at higher frequencies (see Fig.\,\ref{fig2}),
    the normalized Fourier amplitude is shown in Fig.\,\ref{fig3} rather than the original value.
In practice, at any given frequency, the Fourier amplitude at an arbitrary height
    is normalized by the corresponding
    value at 965\arcsec.
The first impression from Fig.\,\ref{fig3} is that
    the overall behavior of the amplitude distribution is similar in the three examined regions.
In particular, the distribution is similar for the plume and inter-plume regions,
    suggesting that the differences in the physical parameters therein
    do not qualitatively change the dependence of the Fourier amplitude
    on height and frequency.
Furthermore, the spatial damping does not show a monotonic frequency dependence.
 It clearly shows that this damping tends to increase with frequency first in lower frequencies
       and then decrease in higher frequencies.
    This dependency is clearly seen
in the results of the entire coronal hole (top row in Fig.\,\ref{fig3}) where the frequency-dependence changes its behavior around 0.7 mHz.
A similar dependency of the damping in plume and inter-plume regions can also be seen, although not very clear.

The frequency dependence of the spatial damping of intensity disturbances is further examined in Fig.~\ref{fig4},
    where the spatial profiles of the normalized Fourier amplitude at a number
    of different frequencies are displayed as labeled.
One sees that regardless of the region examined, the spatial damping of intensity disturbances tends to be stronger
    at lower frequencies.
However, it turns out that deducing a damping length is not straightforward, since
    the spatial profiles at the chosen frequencies do not possess an exponential height dependence
    (note that a globally exponential behavior would yield a straight line given that the amplitude is plotted on a log scale).
Nevertheless, it is possible to deduce a local damping factor based on a linear fit to a few (5 in this study) data points centering at a given height by assuming that the amplitude follows an exponential damping law locally. The damping factor is given by the local slope obtained from the linear fit operated as above. The damping factor quantifies how fast the amplitude of a disturbance decreases when it propagates in the solar atmosphere.

\par
Figure~\ref{fig5} presents the height variation of the damping factor for the same set of frequencies.
    The damping profiles seems to be close to Gaussian below 980\arcsec, and then the damping factor nearly settles to a constant beyond 980\arcsec,
    suggesting that the spatial damping becomes an exponential one.
However, the damping factor below 980\arcsec\ shows some considerable height dependence.
This indicates that the mechanisms for spatially damping the intensity disturbances below 980\arcsec
    are different from those beyond.
Despite this complication, the damping factor at a given height
    tends to decrease with frequency.
Note that the two-step spatial damping and the inverse correlation between frequency and the damping factor
    were also noticed by \citet[][their Figs. 7\&8]{2014A&A...568A..96G}.
These two features, the frequency dependence of damping factor in particular,
    seem difficult to understand if intensity disturbances are a continuous spectrum of slow waves emitted by, e.g.,
    a broadband source at some fixed location.
The reason is that, with the sound speed given,
    the wavelength becomes shorter with increasing frequency
    and consequently slow waves are expected to experience stronger damping, provided that
    ion viscosity or electron heat conduction are the primary damping mechanisms \citep[see also][]{2014ApJ...789..118K}.

\subsection{Power spectra of the intensity disturbances}
More insights can be gained by examining Fig.\,\ref{fig6}, where the power spectra of the intensity disturbances at a number of heights
    are shown for the three examined regions imaged in both 171\,\AA\ and 193\,\AA\ channels.
    The power spectra at a given height in a region are obtained from averaging the whole area of the region parallel to the limb (see Section\,\ref{sect:obs}) and 9 pixels around the given height.
Fittings are given for a power spectrum of each pixel in each region at a given height and a given observing passband
        to reveal the power-law behavior of the spectra.
The average and standard deviation ($\sigma$) of the power-law indices at a given height and a given passband are given in Fig.\,\ref{fig6}.
Comparing different passbands, one can see that the power-law index ($\alpha$) is slightly larger in magnitude
     in the hotter channel (AIA 193\,\AA). However, the difference is small and cannot be distinguished within one sigma deviation.
While comparing different regions, we found the power-law indices are larger in magnitude in AIA 193\,\AA\ plume region.
One can also see that the power-laws provide a rather satisfactory description for the power spectra
    until a knee is reached at $\sim 2-3$ mHz, beyond which the spectra flattens.
At the lowest height examined (965\arcsec),
    in all three regions, some excess power is present in the frequencies ranging from 0.4\,mHz to 2\,mHz (i.e. 8--40 minutes in period).
As was discussed in \citet{2015ApJ...798....1I}, this excess power may derive from upward propagating disturbances generated in the lower parts of the solar atmosphere.
What is new in Fig.~\ref{fig6} is that the spectra slope beyond the spectral knee
    tends to decrease in magnitude with height.
    This is also at variance with the interpretation
    that intensity disturbances are propagating slow waves generated by a continuous broadband driver.
The reason is, the mechanisms dominating the spatial dependence of the Fourier amplitudes
    either are frequency-insensitive (gravitational stratification and magnetic field divergence)
    or provide a damping efficiency that correlates positively with frequency (ion viscosity and electron heat conduction)~\citep{2014ApJ...789..118K}.

Given a power-law-like behavior both below and beyond the spectral knee, one may speculate that
    turbulence may be at work that transfers energy from lower to higher frequencies, thereby
    replenishing the power at higher frequencies and consequently increasing the effective
    damping length.
However, this is unlikely the case for the following two reasons.
First, most of the obtained spectral slopes, except the ones obtained in the plume region observed by the AIA 193\,\AA,
    range from $-1.1$ to $-1.4$ (the uncertainties of these indices are rather substantial though). These nominal values agree with neither the value expected in a low-$\beta$ environment for isotropic hydrodynamic turbulence ($-5/3$)
    nor the ones expected for isotropic ($-3/2$) or anisotropic ($-2$ for the parallel wavenumber spectra) MHD turbulence
    \citep{2003LNP...614...56C}.
Second, if the spectral transfer is due to nonlinear interactions among slow waves, the efficiency of this transfer
    is expected to be at most proportional to the square of the amplitude of the waves at low frequencies.
However, the amplitude of the intensity disturbances summed over all frequencies amount to only a few percent of the background intensity,
    let alone the amplitude in a particular frequency range.
The consequent nonlinear interaction is therefore expected to be rather inefficient.

So now how to understand the frequency-dependent spatial damping of the intensity disturbances in view of their frequency spectra?
Our tentative interpretation is that, rather than comprising a continuous spectrum of slow waves,
    these spectra may reflect a series of intermittent slow waves where adjacent slow wave pulses
    are rather randomly separated in response to the activities that excite these pulses.
What this means is that not all frequency components in a spectrum can be attributed to waves, and
    consequently, the spatial damping of individual components at various frequencies does not necessarily
    need to be contrasted with the theoretical expectations of slow wave damping.
For illustration purposes, consider the toy model where some activity repeats, say, every 10 minutes in the transition region or chromosphere.
Each instance of this activity generates a velocity pulse that impinges on the corona, leading to the eventual development
    of a slow wave pulse~\citep[e.g.,][]{2013ApJ...775L..23W}.
Now that slow waves are dispersion-less, the intensity time series at some coronal height will consist of
    a series of wave pulses with a temporal spacing of 10 minutes.
This 10-minute periodicity will make it into the frequency spectrum when the time series is Fourier-analyzed.
However, it does not correspond to any wave signal, and the spatial variation of the spectral component in a frequency range
    around this periodicity does not need to agree or disagree with wave theory.
In reality, the temporal spacing between two instances of activities lower down should be much more irregular than
    being uniform.
The corresponding spectral components, which themselves do not correspond to waves, are expected to appear in the
    frequency spectra of the intensity disturbances as well.

Figure~\ref{fig7} provides an observational test of the above-mentioned picture, where
    the frequency spectrum of the repetitive spicular activities measured by AIA304\,\AA\
    is compared with that of the intensity disturbances in the corona (AIA 171\,\AA).
As recently found by \citet{2015ApJ...809L..17J} and \cite{2015arXiv151107354S},
    spicular activities are an important source for generating coronal intensity disturbances.
One finds from Fig. \ref{fig7} that the two spectra are indeed well correlated, lending support to the notion that
    the frequency components in the exciters (spicular activities in the present case)
    can contribute to the spectra of coronal intensity disturbances.

%\section{Discussion on the physical nature of the intensity disturbances}
%\label{sect:dis}
%The {\bf damping}s and power spectra presented above lead us to think what is the physical nature of the intensity disturbances in the polar coronal hole. It challenges the slow magnetoacoustic wave scenario since the {\bf damping} of the intensity disturbances in low frequency is stronger than that in higher frequency. ......

%\par
%The power spectra of the intensity disturbances show an additional enhanced component in a period range of 8--40 minutes at lower height, which suggests of a possible contribution of spicular activities.

\section{Summary}
\label{sect:sum}
We have examined the damping and power spectra of the intensity disturbances in a polar coronal hole.
The damping of the intensity disturbances at low frequencies is stronger than that at higher frequencies.
The damping of the intensity disturbances becomes exponential only above 980\arcsec, suggesting that
    the intensity disturbances may experience a two-step damping rather than being damping by a single mechanism operating
    at all heights.
We found that the power spectra of the intensity disturbances show power-laws with roughly the same index in different regions
    in a chosen AIA passband.
 While comparing different passbands, one can see that the power-law index of the power spectra is slightly larger in magnitude
    in the hotter channel (AIA 193\,\AA). However, the difference is small and cannot be distinguished within one sigma deviation.
Comparing different regions, we found it is larger in magnitude in AIA 193\,\AA\ plume region.
An additional enhanced component is present in the power spectra in a period range of 8--40 minutes at lower heights,
    which suggests a possible contribution of upward propagating waves from lower parts of the solar atmosphere.
The observed damping and power spectra of the intensity disturbances are found to be at variance with the interpretation that
    coronal intensity disturbances are a continuous spectrum of slow waves excited by broadband drivers.
Instead, coronal intensity disturbances may consist of randomly separated slow wave pulses with a substantial number of their frequency components
    bearing signatures of the exciters.
One important exciter is attributed to spicular activities as seen in the AIA 304\,\AA\ channel.

\normalem
\begin{acknowledgements}
We thank the anonymous referee for his/her critical reading and constructive comments.
This research is supported by the China 973 program 2012CB825601, the National Natural Science Foundation of China under contracts: 41404135 (Z.H.), 41274178 and 41474150 (L.X. \& Z.H.), and 41174154, 41274176 and 41474149 (B.L.), the Shandong provincial Natural Science Foundation ZR2014DQ006 (Z.H.). AIA data is courtesy of SDO (NASA). We thank JSOC for providing downlinks of the SDO data.
\end{acknowledgements}

\bibliographystyle{raa}
\bibliography{references}

\begin{thebibliography}{29}
\providecommand\natexlab[1]{#1}
\providecommand\JournalTitle[1]{#1}

\bibitem[{Banerjee} {et~al.}(2009)]{2009A&A...499L..29B}
{Banerjee}, D., {Teriaca}, L., {Gupta}, G.~R., {et~al.} 2009, \aap, 499, L29

\bibitem[{Cho} {et~al.}(2003)]{2003LNP...614...56C}
{Cho}, J., {Lazarian}, A., \& {Vishniac}, E.~T. 2003, in Lecture Notes in
  Physics, Berlin Springer Verlag, Vol. 614, Turbulence and Magnetic Fields in
  Astrophysics, ed. E.~{Falgarone} \& T.~{Passot}, 56

\bibitem[{De Moortel} {et~al.}(2015)]{2015SoPh..290..399D}
{De Moortel}, I., {Antolin}, P., \& {Van Doorsselaere}, T. 2015, \solphys, 290,
  399

\bibitem[{De Moortel} \& {Hood}(2003)]{2003A&A...408..755D}
{De Moortel}, I., \& {Hood}, A.~W. 2003, \aap, 408, 755

\bibitem[{De Moortel} \& {Hood}(2004)]{2004A&A...415..705D}
{De Moortel}, I., \& {Hood}, A.~W. 2004, \aap, 415, 705

\bibitem[{De Moortel} {et~al.}(2004)]{2004A&A...425..741D}
{De Moortel}, I., {Hood}, A.~W., {Gerrard}, C.~L., \& {Brooks}, S.~J. 2004,
  \aap, 425, 741

\bibitem[{DeForest} \& {Gurman}(1998)]{1998ApJ...501L.217D}
{DeForest}, C.~E., \& {Gurman}, J.~B. 1998, \apjl, 501, L217

\bibitem[{Fang} {et~al.}(2015)]{2015ApJ...813...33F}
{Fang}, X., {Yuan}, D., {Van Doorsselaere}, T., {Keppens}, R., \& {Xia}, C.
  2015, \apj, 813, 33

\bibitem[{Gupta}(2014)]{2014A&A...568A..96G}
{Gupta}, G.~R. 2014, \aap, 568, A96

\bibitem[{Gupta} {et~al.}(2012)]{2012A&A...546A..93G}
{Gupta}, G.~R., {Teriaca}, L., {Marsch}, E., {Solanki}, S.~K., \& {Banerjee},
  D. 2012, \aap, 546, A93

\bibitem[{Ireland} {et~al.}(2015)]{2015ApJ...798....1I}
{Ireland}, J., {McAteer}, R.~T.~J., \& {Inglis}, A.~R. 2015, \apj, 798, 1

\bibitem[{Jiao} {et~al.}(2015)]{2015ApJ...809L..17J}
{Jiao}, F., {Xia}, L., {Li}, B., {et~al.} 2015, \apjl, 809, L17

\bibitem[{Krishna Prasad} {et~al.}(2011)]{2011A&A...528L...4K}
{Krishna Prasad}, S., {Banerjee}, D., \& {Gupta}, G.~R. 2011, \aap, 528, L4

\bibitem[{Krishna Prasad} {et~al.}(2014)]{2014ApJ...789..118K}
{Krishna Prasad}, S., {Banerjee}, D., \& {Van Doorsselaere}, T. 2014, \apj,
  789, 118

\bibitem[{Krishna Prasad} {et~al.}(2012)]{2012A&A...546A..50K}
{Krishna Prasad}, S., {Banerjee}, D., {Van Doorsselaere}, T., \& {Singh}, J.
  2012, \aap, 546, A50

\bibitem[{Lemen} {et~al.}(2012)]{2012SoPh..275...17L}
{Lemen}, J.~R., {Title}, A.~M., {Akin}, D.~J., {et~al.} 2012, \solphys, 275, 17

\bibitem[{McIntosh} {et~al.}(2010)]{2010A&A...510L...2M}
{McIntosh}, S.~W., {Innes}, D.~E., {de Pontieu}, B., \& {Leamon}, R.~J. 2010,
  \aap, 510, L2

\bibitem[{Ofman} {et~al.}(1999)]{1999ApJ...514..441O}
{Ofman}, L., {Nakariakov}, V.~M., \& {DeForest}, C.~E. 1999, \apj, 514, 441

\bibitem[{Ofman} {et~al.}(2000)]{2000ApJ...533.1071O}
{Ofman}, L., {Nakariakov}, V.~M., \& {Sehgal}, N. 2000, \apj, 533, 1071

\bibitem[{Ofman} {et~al.}(1997)]{1997ApJ...491L.111O}
{Ofman}, L., {Romoli}, M., {Poletto}, G., {Noci}, G., \& {Kohl}, J.~L. 1997,
  \apjl, 491, L111

\bibitem[{Ofman} {et~al.}(2012)]{2012ApJ...754..111O}
{Ofman}, L., {Wang}, T.~J., \& {Davila}, J.~M. 2012, \apj, 754, 111

\bibitem[{Samanta} {et~al.}(2015)]{2015arXiv151107354S}
{Samanta}, T., {Pant}, V., \& {Banerjee}, D. 2015, arXiv:1511.07354

\bibitem[Su(2014)]{su2014b}
Su, J.~T. 2014, \apj, 793, 117

\bibitem[Su {et~al.}(2014)]{su2014a}
Su, J.~T., Liu, Y., Shen, Y.~D., \& Priya, T.~G. 2014, \apj, 790, 150

\bibitem[{Tian} {et~al.}(2011)]{2011ApJ...727L..37T}
{Tian}, H., {McIntosh}, S.~W., \& {De Pontieu}, B. 2011, \apjl, 727, L37

\bibitem[{Verwichte} {et~al.}(2008)]{2008ApJ...685.1286V}
{Verwichte}, E., {Haynes}, M., {Arber}, T.~D., \& {Brady}, C.~S. 2008, \apj,
  685, 1286

\bibitem[{Wang} {et~al.}(2013)]{2013ApJ...775L..23W}
{Wang}, T., {Ofman}, L., \& {Davila}, J.~M. 2013, \apjl, 775, L23

\bibitem[{Withbroe}(1983)]{1983SoPh...89...77W}
{Withbroe}, G.~L. 1983, \solphys, 89, 77

\bibitem[{Zhang} {et~al.}(2015)]{2015A&A...581A..78Z}
{Zhang}, Y., {Zhang}, J., {Wang}, J., \& {Nakariakov}, V.~M. 2015, \aap, 581,
  A78

\end{thebibliography}

\end{document}